# Second-harmonic magnetic response characterizing magnetite-based colloid

V.A. Ryzhov[a], I.A. Kiselev[a], O.P. Smirnov[a], Yu.P. Chernenkov[a], V.V. Deriglazov[a,*],
Ya.Yu. Marchenko[b], L.Y. Yakovleva[b], B.P. Nikolaev[b], Yu.V. Bogachev[c]

[a] *Petersburg Nuclear Physics Institute named by B.P. Konstantinov of National Research Centre "Kurchatov Institute", 1 Orlova roscha mcr., 188300 Gatchina, Leningrad Region, Russia*
[b] *Research Institute of Highly Pure Biopreparations, 7 Pudozhskaya st, 197110 St-Petersburg, Russia*
[c] *St-Petersburg State Electrotechnical University LETI, 5 Prof. Popov st, 197376 St-Petersburg, Russia*

A B S T R A C T

Nonlinear second-harmonic magnetic response (M2) was used to characterize an aqueous colloidal solution of dextran-coated magnetite ($Fe_3O_4$) nanoparticles. Data analysis with the formalism based on Gilbert-Landau-Lifshitz equation for stochastic dynamics of superparamagnetic (SP) particles ensured extensive quantifying of the system via a set of magnetic and magnetodynamic parameters, such as the mean magnetic moment, the damping constant, the longitudinal relaxation time, the magnetic anisotropy field and energy, and others. Combined with transmission electron microscopy and dynamic light scattering, M2 technique allowed obtaining additional parameters, viz., the dextran-coating thickness and the interparticle magnetic dipolar energy. Aggregated colloidal nanoparticles were shown to be magnetically correlated inside the aggregate due to magnetic dipole-dipole (d-d) coupling within the correlation radius ~50 nm. With the d-d coupling account, the volume distribution of the aggregates recovered from M2 measurements is well consistent with electron microscopy results. From electron magnetic resonance, abrupt change of SP dynamics with increasing external magnetic field was observed and explained. The presented study exemplifies a novel M2-based procedure of comprehensive quantitative characterization applicable for a wide variety of SP systems.

*Keywords:* Magnetic nanoparticles, Colloids, Superparamagnetism, Nonlinear magnetic response, Electron magnetic resonance

## 1. Introduction

Magnetic nanoparticles (MNPs) are widely adopted in technical, environmental and biomedical areas [1-9]. Such MNPs as superparamagnetic iron-oxide nanoparticles (SPIONs) are widely used in disease diagnostics as contrast agents in magnetic resonance imaging and in tumor treatment including hyperthermia and drug delivery. Magnetite-based SPIONs are among the most demanded MNPs due to their biocompatibility and zero coercivity which helps prevent aggregation *in vivo* [7, 8]. For many applications, SPIONs are dispersed in liquids forming suspensions or colloidal solutions to be injected or consumed in some other way. To stabilize the SPIONs in the carrier liquid and to prevent toxicity and oxidation, they are coated with specific shells of organic compounds [2, 8], one of which being dextran, a carbohydrate incorporating the polymers of glucose.

However, depending on the solvent, the concentration, the coating fraction and other factors, the dispersed nanoparticles may aggregate [7, 10]. This tendency hampers colloidal stability needed in biomedicine and modifies relevant magnetic features. Meanwhile, information on the size distribution of aggregates obtained by different techniques is rather poor and contradictory. Thus, electron microscopy needs drying the suspension what can modify the size distribution. Another example is dynamic light scattering (DLS) which yields the hydrodynamic diameter, hard to be related to the true geometric size. Magnetic techniques may somewhat underestimate the aggregate size due to the fact that, inside an aggregate, SPIONs are coupled by dipole-dipole interaction resulting in hardly accountable finite-radius

---

correlations of MNP magnetic moments [11]. Also, the study of magnetization dynamics of SPION containing systems faces considerable difficulties in conventional magnetic measurements. The details of self-organization of the colloidal solution, as well as its magnetic characteristics, still poorly known, are topics of the present study.

To fabricate ferrofluids with definite attributes, a set of techniques is used for their characterization. The conventional toolkit includes steady-field magnetometry registering also magnetic hysteresis, ac susceptibility measurements, X-ray diffraction (XRD), and transmission electron microscopy (TEM). Additional information can be obtained from magnetorelaxometry [12], Raman spectroscopy, DLS also known as photon correlation spectroscopy, magnetoresistivity measurements [13] and some others. As magnetic properties of SPION containing systems, especially, the magnetization dynamics, are the most meaningful for certain applications (e.g. hyperthermia), the instruments directly aiming at this domain are of particular significance. The list of such techniques permanently expands indicating topicality of the issue. One of the latest is the nonlinear ac Faraday rotation applied to characterization of magnetite-based SPION aqueous suspensions [14].

Here, we employ a technique involving nonlinear magnetic response on the second harmonic (M2) in the longitudinal geometry of ac- and dc magnetic fields which is perfectly well available to study SP systems. First, in the megahertz frequency range, particles with the relevant magnetic moments ~$10^3$ – $10^6$ $\mu_B$ generate an appreciable nonlinear response with pronounced extrema in small magnetic fields of the order ~10 – 100 Oe. Second, on the same conditions, real and imaginary parts of the response are comparable, hence, providing, jointly, a great informational content of the measurements. Third, this technique is highly sensitive to the SP magnetization dynamics.

Capabilities of the M2 technique were repeatedly demonstrated in elaborating a set of condensed matter [15-17] and biophysical [18] issues. In particular, its efficiency has shown itself in studying magneto-electronic phase separation in 3$d$ oxides [19-21]. As a result, a lot of novel information was obtained on the emerging system of ferromagnetic nanoclusters from *qualitative* analysis of the raw data. In the same way, a biodistribution of magnetite nanoparticles in tissues of rodents in studies of the brain tumor targeting was evaluated by this technique [18]. In the present study, this experimental resource was supplemented with the recently elaborated rigorous formalism based on Gilbert-Landau-Lifshitz (GLL) equation for stochastic dynamics of SP particles [22-24]. Treating M2 experimental data with this formalism enables to extract a full scope of magnetic and magnetodynamic parameters characterizing such systems. Here, the GLL data-treatment formalism was employed to describe *quantitatively* an aqueous colloidal solution of magnetite-dextran SPIONs. Such an object may be considered as a trial prototype of more complicated magnetic sols relevant for perspective nanomedical applications.

However, the M2 technique alone turned out to be insufficient for total unambiguous characterization of this rather complicated system. Therefore, conventional nonmagnetic techniques such as XRD, TEM and DLS have been additionally enabled to verify and add to M2 data. It concerns, mainly, geometrical parameters, such as the mean particle volume and the volume distribution width, which depend on the nonmagnetic component of the system invisible by magnetic probes. Thus, particular attention was paid to mutual consistency of the data obtained by all these techniques. Jointly, they facilitate correct interpretation of M2 data and yield additional information unavailable from the M2 technique alone.

This approach is intended to be further applied to monitor functionalized SPIONs accumulated in various organs and tissues of experimental animals. It will provide information on the coupling of the SPIONs with specific sites of different tissue cells, including tumors, and on aggregation of the SPIONs. Particular data obtained in the present study will be useful as a reference point.

EMR measurements were also performed. The resonance spectrum was fitted with the same GLL formalism and explained consistently with M2 and TEM observations.

In Section 2, preparation of the colloidal solution is briefly described and its attestation by XRD is presented. In Section 3, TEM and DLS data are analyzed and compared. In the main Section 4, a concise description of the M2 technique and the measurement conditions are offered. The data treatment formalism is also traced, with the reference on more detailed description. The parameters characterizing the nanoparticle system are analyzed and the structural, magnetic and magnetodynamic properties are discussed. In Section 5, EMR data are presented and analyzed involving information obtained from M2 and TEM measurements. The conclusion is presented in Section 6.

## 2. Sample preparation and attestation

SPIONs were prepared from solutions of the iron salts $FeSO_4$ and $FeCl_3$ at the ratio of ion concentrations $Fe^{+2}/Fe^{+3} = 1/2$ by co-precipitation in alkaline media at the temperature 80°C under the inert gas $N_2$ [25]. To ensure disaggregation, low molecular weight dextran (MW 10 kDa, Sigma) supplemented by CsCl was added to the dispersion of magnetic nanoparticles in the process of sonication during 15 min at the frequency 22.3 kHz. Precipitation was initiated by continual titration with $NH_4OH$ solution under stirring in the 100 mL reactor. The stock dispersion was collected by the permanent Nd magnet and then washed and centrifuged into fractions. A fine fraction of the SPIONs was treated by dialysis and stored at 4°C before the experiments. The Fe content was assayed with the thiocyanate probe by measuring the light absorption at the wavelength 480 nm.

The SPIONs structure and composition were examined by XRD at DRON-3M diffractometer. Fig. 1 presents the XRD intensity as a function of the diffraction angle measured at the temperature 290 K.

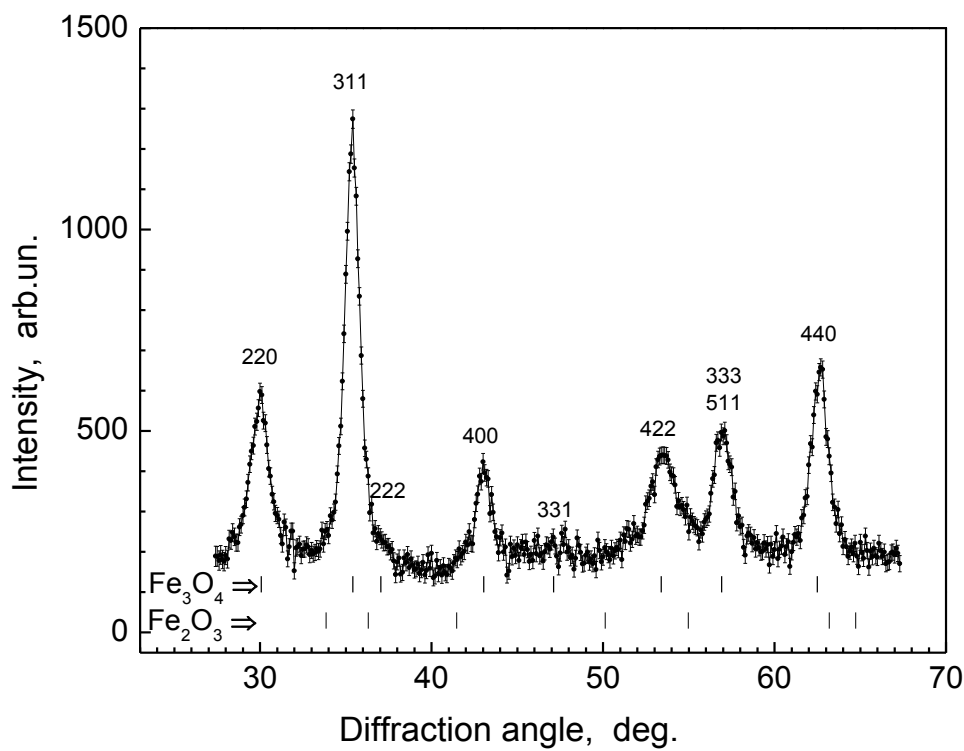

**Fig. 1.** X-ray diffraction intensity vs. diffraction angle for Fe-oxide nanocrystals. Marks at the bottom indicate nominal reflections: the upper and the lower sets are for magnetite and hematite, respectively.

To evaluate a size of the nanoparticle crystallinity region, precise treatment of the XRD pattern was performed with account of the instrumental resolution and a doublet structure of Cu $K_\alpha$ line. The diffraction peaks broaden, mainly, due to a finite size of the coherent scattering region and internal stress in the sample. Williamson–Hall approach [26] clearly differentiates between the size-induced and strain-induced peak broadening by considering the peak width as a function of angle:

$$\beta_{hkl} \cos\theta = \frac{k\lambda}{d} + 4\varepsilon \sin\theta$$

where $\beta_{hkl}$ is the instrument-corrected breadth (full width at half maximum) of *hkl*-reflection located at the angle $2\theta$, *d* is the crystallite size, $k \approx 0.9$, $\lambda = 1.54$ Å is the wavelength of Cu K$_{\alpha 1}$ radiation and $\varepsilon$ is the strain-induced broadening arising from crystal imperfections and distortions. From the meaningful peaks of Fig. 1, the mean size of the crystallinity region in Fe$_3$O$_4$ nanoparticles was found to be $\bar{d} =$ 8.7(1.3) nm, the strain-induced broadening being small.

For further study, the SPIONs were coated with dextran and dispersed in water to form a colloidal solution [25].

## 3. Probing by transmission electron microscopy and dynamic light scattering

*3.1. Transmission electron microscopy*

Fig. 2 (upper panel) presents a fragment of the TEM micrograph obtained at the microscope JEM-100C (Jeol, Japan) for the freeze-dried colloid on the glass substrate.

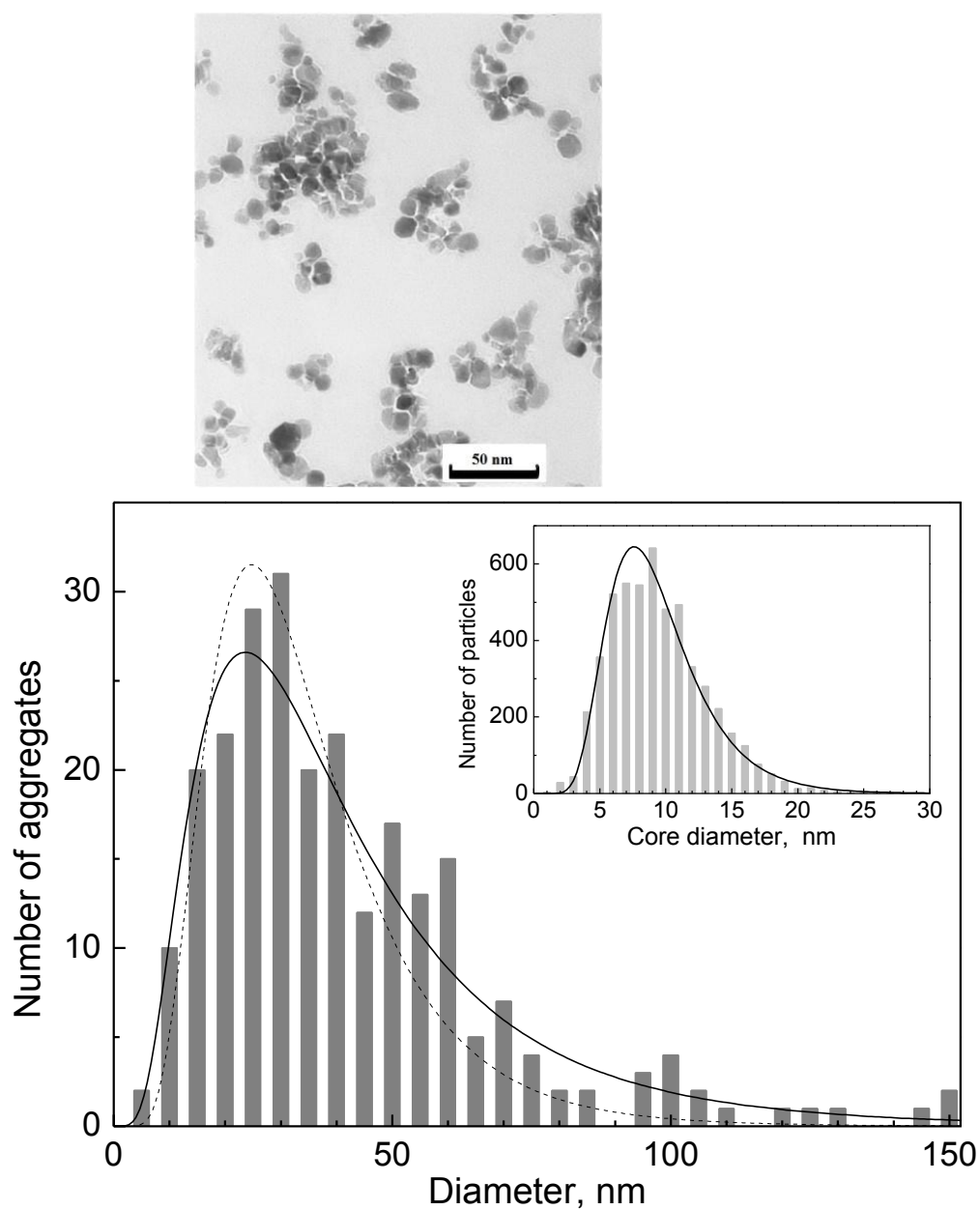

**Fig. 2.** Upper panel: TEM image of freeze-dried colloid; lower panel: measured diameter distribution of aggregates (histogram) and its best fit with lognormal distribution (solid curve); inset: the same for distribution of magnetite cores of constituting nanoparticles. Dashed curve is lognormal diameter distribution recovered from DLS histogram of Fig. 3 (see Subsection 3.2).

Nanoparticles in the form of granules ~10 nm in diameter make up aggregates of different size and irregular form. The micrograph contrast is caused totally by magnetite cores while dextran shells around the cores are only slightly discernible as thin gaps between dark spots. On the lower panel, a size distribution of the aggregates is shown as a function of a diameter of the effective sphere approximating an aggregate. In the inset, the same distribution is presented for magnetite cores of nanoparticles, building units of the aggregates. Both the systems are expectedly well fitted by the lognormal distribution (solid lines):

$$f(x) = \frac{1}{\sqrt{2\pi}\sigma x} \exp\left(-\frac{1}{2\sigma^2}\ln^2\frac{x}{x_0}\right) \qquad (1)$$

with the mean $x$-value $\bar{x} = x_0\exp(\sigma^2/2)$ and the variance $x_0^2\exp\sigma^2(\exp\sigma^2 - 1)$. The diameter distribution parameters are given in Table 1.

**Table 1**
Parameters of diameter lognormal distribution for nanoparticle magnetite cores (left column) and nanoparticle aggregates (right column) as obtained from TEM.

|  | Particle cores | Aggregates |
| --- | --- | --- |
| Median, nm | 8.79(5) | 35(2) |
| Mean diameter, nm | 9.46(5) | 43(2) |
| Standard deviation | 0.383(4) | 0.63(4) |
| Variance, nm$^2$ | 14.1(3) | 880(150) |

Within the statistical errors, the mean diameters and the variances coincide with these obtained directly from the histograms.

Note, the crystallinity size extracted from XRD is equal, within the measurement accuracy, to the mean diameter of the magnetite cores evaluated with TEM. Thus, no non-crystal phase was detected in the magnetite fraction.

The lognormal distribution over diameters with the median $D_0$ and the standard deviation $\sigma_D$ means also the lognormal distribution over particle volumes with the median $V_0 = \pi D_0^3/6$ and the standard deviation $\sigma_V = 3\sigma_D$. The mean diameter $\bar{D} = D_0\exp(\sigma_D^2/2)$ and the diameter $\widetilde{D}$ corresponding to the mean volume $\bar{V} = V_0\exp(\sigma_V^2/2)$ via $\bar{V} = \pi\widetilde{D}^3/6$ interrelate as $\widetilde{D} = \bar{D}\exp\sigma_D^2$.

The particle cores are expected to be in the single-domain state which extends up to the size 128 nm [27]. The latter will be confirmed by M2 data presented below.

*3.2. Dynamic light scattering*

The DLS measurements were performed with Zetasizer Nano ZSP (Malvern) equipment for the samples in glass cylindrical cuvettes with the internal diameter 10 mm and the volume 2 mL at the temperature 298 K. The most part of the colloid, 98.5%, was observed in the aggregated state.

Three independent DLS measurements have been performed for this sample. In Fig. 3, a typical histogram of the hydrodynamic diameter distribution is presented. Unlike the microscopy data (Fig. 2), the lognormal distribution (dotted line) insufficiently describes the DLS histogram, with noticeable deviations at the histogram edges. The hydrophilic dextran shell of the particles couples with water by hydrogen bonds creating a solvent layer bound to the aggregate surfaces. A part of water can also penetrate inside an aggregate. Besides, due to a strongly irregular form, the aggregates, when moving, capture some amount of water. So, an effective thickness of the water layer is not proportional to the aggregate size resulting in deviation of the measured hydrodynamic diameter from the lognormal distribution. The effective aggregate diameter can be presented as $D = D_h - 2\Delta$, where $D_h$ is the hydrodynamic diameter as seen from DLS and $\varDelta$ is the effective thickness of the water layer. With the assumption that $D$ conforms to the lognormal distribution and $\varDelta$ is independent of $D$, the fitting was performed for each of the three measurements. For all of them, the fit curves perfectly well lay on the respective histograms as shown for one of them in Fig. 3 (solid line).

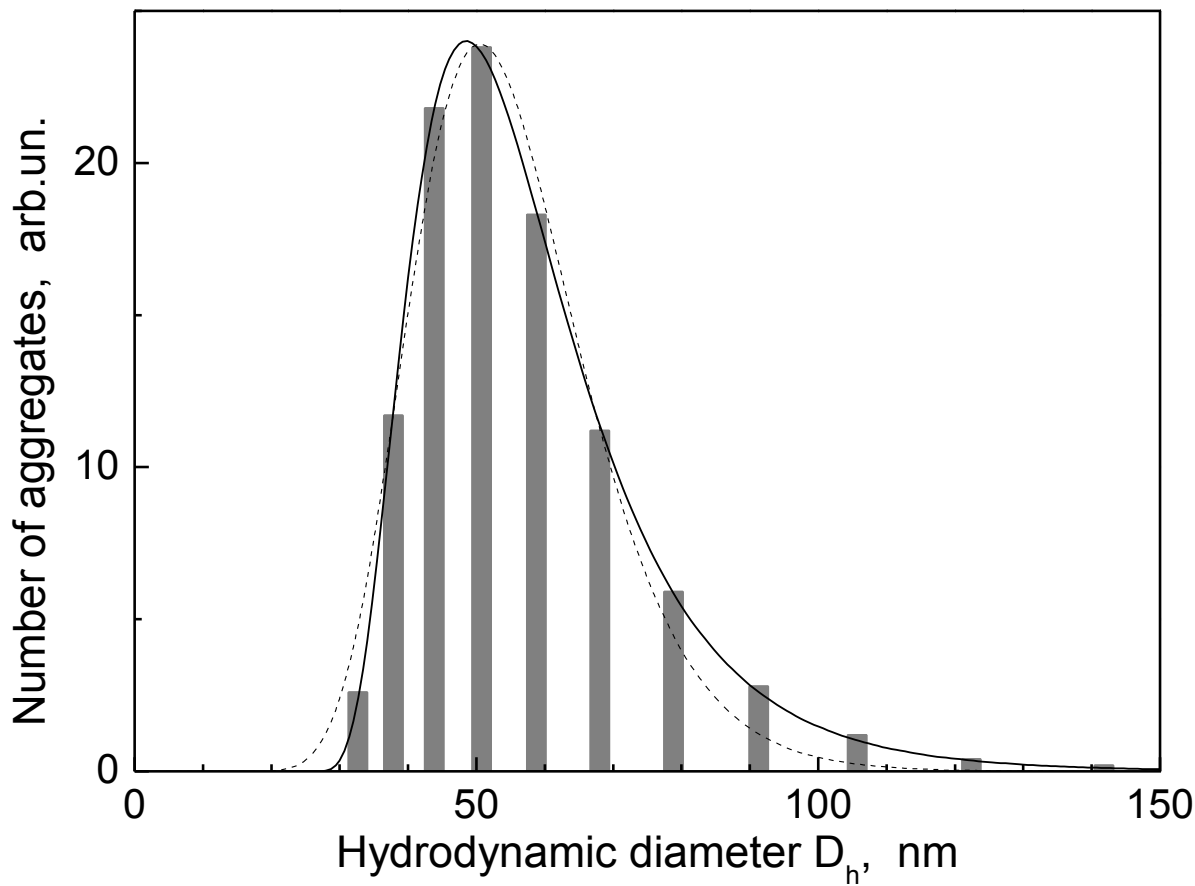

**Fig. 3.** Distribution of aggregates hydrodynamic diameters (histogram) and its best fits: with lognormal distribution without water layer account (dotted line) and with water layer account (solid line).

The parameters averaged over the three measurements with the errors within the reproducibility (the confidence level 0.95) are presented in Table 2.

**Table 2**
Parameters of diameter distribution for nanoparticle aggregates obtained from DLS: without account of water layer (left column) and with the water layer account (right column).

|  | Water layer disregarded | Water layer accounted |
|---|---|---|
| Median, nm | 50(7) | 28(5) |
| Mean hydrodynamic diameter, nm | 52(7) | 44(7) |
| Mean geometrical diameter, nm | - | 32(5) |
| Standard deviation | 0.25(1) | 0.51(5) |
| Water layer thickness, nm | 0 | 5.8(8) |

As seen from Table 2, account of the water layer appreciably reduces the measured mean diameter and increases the distribution width, approaching them to the TEM aggregate parameters (Table 1). However, the DLS mean diameter still somewhat exceeds the microscopy one. To compare the DLS and TEM results in more detail, the lognormal distribution for the effective diameter $D$ was recovered with the fit parameters of the DLS histogram and presented in Fig. 2 by the dashed curve. The curve is seen to fairly well match the TEM histogram peak, thus, demonstrating a good agreement of both the mea-surements. Jointly, they may be accepted as a base to verify the information obtained from the nonlinear magnetic response.

## 4. Nonlinear magnetic response

### 4.1. Measurements

Colloids with two different concentrations, 2 and 0.02 mM(Fe)/L, were examined with the well developed unconventional technique [18, 28, 29] exploiting the second harmonic $M_2$ of nonlinear magnetic response in parallel ac- and dc magnetic fields, $H(t) = H + h_0 \sin \omega t$. The dc field $H$ was scanned back-and-forth symmetrically within ±300 Oe with the round-up cycles 0.125 – 4 s and with high representativity of 2048 $H$-points in each scan. The amplitude $h_0 = 13.8$ Oe of the ac field with the frequency $f = \omega/2\pi = 15.7$ MHz ensured the condition $M_2 \propto h_0^2$. This nonrigid requirement enabled to directly visualize undistorted $H$-dependence of the second-order susceptibility and somewhat favored reliability of the data treatment. Both, real and imaginary, components of the signal were simultaneously recorded as functions of the dc field at the temperature region 273 – 297K with the temperature stabilization ±0.1 K.

In Fig. 4, the $H$-field direct and reverse scans for real and imaginary parts of the $M_2$ signal from the sample with the concentration 0.02 mM(Fe)/L at the temperature 297 K and the round-up cycle 0.125 s are presented as a typical example.

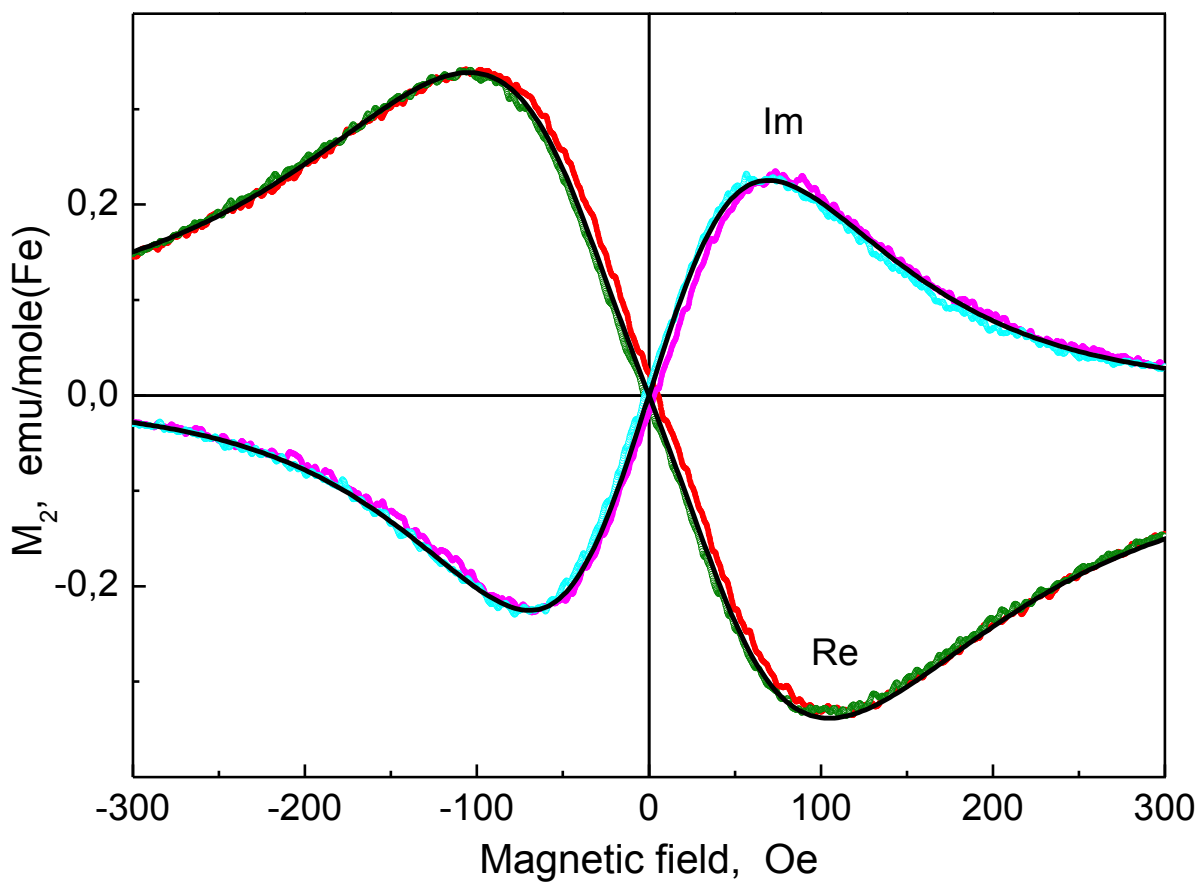

**Fig. 4.** Real and imaginary parts of nonlinear magnetic response as a function of steady magnetic field: filled circles (red and magenta) are direct scan, open circles (green and cyan) are reverse scan and solid curves are simultaneous best fit.

Due to common symmetry requirement, the field dependence of the signal is antisymmetric with respect to zero. The two signal components exhibit opposite signs, pronounced extrema, and only slightly dis-cernible hysteresis, all characteristic for SP behavior. Increase of the round-up cycle up to 4 s leads to complete disappearance of the hysteresis indicating its dynamical character. The latter is a characteristic feature of single-domain MNPs [30]. The spectra for all the temperatures appeared to be quite similar manifesting the absence of any temperature evolution of the colloidal system in this region. Similarly, no distinction was found between the spectra for the two concentrations evidencing the absence of noticeable coupling between components of the SP system in this concentration range. Increasing the scan cycle duration up to 4 s expectedly results only in complete elimination of the hysteresis with no effect on the other features, thus, evidencing a dynamical character of the hysteresis inherent to single domain MNPs [30]. Anticipating the results, the similarity of all spectra gives rise to one and the same set of parameters, within the errors, characterizing the system under study.

*4.2. Data treatment*

Treatment of the obtained M2 experimental data was carried out following the formalism elaborated recently by Coffey and colleagues [22-24]. Real and imaginary components of the measured M2 response were simultaneously fitted with the model function containing the stationary solution of the Fokker-Planck equation for the SP magnetic moment. In spherical coordinates, it reads [31]:

$$2\tau_N \frac{\partial W}{\partial t} = -\frac{1}{\sin\vartheta}\left[\frac{\partial}{\partial \vartheta}(\sin\vartheta \tilde{J}_\vartheta) + \frac{\partial}{\partial \varphi}(\tilde{J}_\varphi)\right] \quad (2a)$$

with

$$\tilde{J}_\vartheta = -\left[\beta\left(\frac{\partial \mathcal{H}}{\partial \vartheta} - \frac{1}{\alpha}\frac{1}{\sin\vartheta}\frac{\partial \mathcal{H}}{\partial \varphi}\right)W + \frac{\partial W}{\partial \vartheta}\right], \quad (2b)$$

$$\tilde{J}_\varphi = -\left[\beta\left(\frac{1}{\alpha}\frac{\partial \mathcal{H}}{\partial \vartheta} + \frac{1}{\sin\vartheta}\frac{\partial \mathcal{H}}{\partial \varphi}\right)W + \frac{1}{\sin\vartheta}\frac{\partial W}{\partial \varphi}\right]. \quad (2c)$$

Here, $W$ is the nonequilibrium probability-density function for directions of the particle magnetic moment, the dimensionless constant $\alpha$ is proportional to the damping factor in the dissipation term of the GLL stochastic equation [31], the time scale $\tau_N \propto \alpha + \alpha^{-1}$ is the Néel time in the Gilbert form, and $\beta = 1/k_B T$ ($k_B$ is the Boltzmann constant). The magnetic potential $\mathcal{H}$ is a sum of the *uniaxial* anisotropy energy and the energy of the magnetic moment in the external magnetic field $H(t)$:

$$\mathcal{H} = -\frac{K_a V}{m^2}(\mathbf{mn})^2 - \mathbf{mH}$$

where $K_a$ is the anisotropy constant, $V$ and $\mathbf{m}$ are the particle volume and magnetic moment, respectively, and the unit vector $\mathbf{n}$ is the anisotropy axis direction. The terms with $1/\alpha$ and the rest terms in Eqs. (2b) and (2c) describe precession and thermal diffusion, respectively.

An analytical solution of the Eq. (2a) for the present case of arbitrary magnetic field direction is absent enforcing to solve the problem numerically. By expanding $W(t)$ in the series on spherical harmonics

$$W(t,\vartheta,\varphi) = \sum_{lm} c_{lm}(t) Y_{lm}(\vartheta,\varphi) \quad (3)$$

and in the Fourier series, the Eq. (2a) is reduced to a linear set of equations which, in turn, can be expressed as a continuous-fraction matrix relation:

$$\mathbf{S}_n = -[\mathbf{Q}_n + \mathbf{Q}^+\mathbf{S}_{n+1}\mathbf{Q}_{n+1}]^{-1} \quad (4)$$

where the matrices $\mathbf{Q}_n$, $\mathbf{Q}^+$, and $\mathbf{Q}_{n+1}$ are composed of the spherical harmonics indices, the direction cosines of the ac- and dc magnetic fields, as well as the parameters entering Eqs. (2a), (2b) and (2c) and the magnetic potential $\mathcal{H}$. As a result, the (normalized) *k*-harmonic of the magnetic moment in the direction of the ac field is expressed as:

$$m_k(\omega) = \sqrt{\frac{4\pi}{3}}\left[\gamma_3' c_{10}^k(\omega) + \frac{(\gamma_1' + i\gamma_2')c_{1-1}^k(\omega) - (\gamma_1' - i\gamma_2')c_{11}^k(\omega)}{\sqrt{2}}\right] \quad (5)$$

where $\gamma_{1,2,3}'$ are the direction cosines of the ac magnetic field while $c_{ij}^k(\omega)$ are Fourier transforms of the expansion coefficients $c_{lm}$ of Eq. (3) and compose the column vector $\mathbf{C}$ proportional to the solution $\mathbf{S}_1$ of the continuous-fraction Eq. (4).

The fit function is a convolution of $m_k(\omega)$ for $k = 2$ with the lognormal magnetic moment distribution:

$$M_2(H) = M_s \int f_M m_2(\omega, h_0, H) dM \quad (6)$$

where $M_s$ is the saturation magnetization of the SP system. The distribution $f_M$ corresponds to the distribution over *volumes* if all particles are magnetically homogeneous with the same, size independent, magnetization.

The solution accuracy is determined by the number of equations in the system, i.e. retained terms in the Fourier and spherical harmonics expansions which, in turn, specify the matrices sizes. The former number ±3 and the latter one 7 ensure a sufficient accuracy. Computation of the fit function (6) at each experimental point implies multiple implementation of the procedure for solution of Eq. (4) (~ $10^2$ per *H*-field point per iterate) while the CPU time rapidly increases with the matrix size. Thus, the total data treatment is essentially time consuming and feasible only at powerful computer clusters. The results of this work were obtained using computational resources of Peter the Great Saint-Petersburg Polytechnic University Supercomputing Center (http://www.spbstu.ru).

## 4.3. Results

### 4.3.1. Effect of magnetic correlations inside aggregates

The fit variable parameters relate to the magnetite fraction of the colloid and include: (1) the saturation magnetization of the SP system per mole of Fe $M_s$, (2) the $f_M$ distribution parameters, viz., the median magnetic moment $M_0$ and (3) the distribution width $\sigma_M$, (4) the mean anisotropy energy $E_a$, (5) the angle $\Psi$ between the anisotropy axis and magnetic field, and (6) the damping constant $\alpha$. Also, the possible backgrounds linearly depending on the steady field $H$ were fitted for real and imaginary parts of the signal. Some additional quantities can be derived from the fit parameters, viz. (i) the anisotropy field $H_a = E_a/\bar{M}$ with the mean magnetic moment $\bar{M} = M_0 \exp(\sigma_M^2/2)$, (ii) the saturation magnetization per Fe ion $\langle\mu\rangle = M_s/N_A$ where $N_A$ is the Avogadro number, (iii) the number of Fe ions corresponding to $\bar{M}$, $\bar{N} = \bar{M}/\langle\mu\rangle$, (iv) the mean volume $\bar{V} = v_0 \bar{N}$ where $v_0 \cong 0.02467$ nm$^3$ is the volume per Fe ion in magnetite obtained from XRD, (v) the mean diameter $\tilde{D}$ corresponding to the mean volume $\bar{V}$, and the mean value $\bar{D} = \tilde{D}\exp(-\sigma_V^2/9)$ of the *diameter* lognormal distribution, (vi) the Néel time $\bar{\tau}_N = \bar{M}\beta(\alpha + \alpha^{-1})/2\gamma$ where $\gamma$ is the gyromagnetic ratio and (vii) the zero-field longitudinal relaxation time $\tau_\parallel$. The latter quantity is proportional to the Neél time, $\tau_\parallel = Q\tau_N$. The proportionality coefficient $Q$ is a monotonously increasing function of $\beta E_a$ inferred and tabulated in the studies [32-34].

Treatments of all the measured spectra yielded nearly the similar sets of parameters. Fig. 4 presents the best fit (solid curves) for the temperature 297 K as a typical case. The parameter values are presented in Table 3 (left column).

The magnetic moment ~$10^5$ $\mu_B$ and the relaxation time ~$10^{-9}$ s are typical for an SP particle. The mean diameter $\bar{D} = 36$ nm being compared to the microscopy diameters from Table 1 indicates that the M2 signal arises from the aggregates rather than from independent SPIONs. This value is smaller than the microscopy one (43 nm) as the former comprises only the magnetically active component of aggregates. Thus, $\bar{D}$ is the *effective* mean diameter corresponding only to the magnetite fraction.

Questionable, however, is the distribution width $\sigma_M$. Its value was expected to fit the triple value of the TEM standard deviation for the aggregates diameter distribution presented in Table 1 (right column), i.e. $\sigma_V = 3\sigma_D = 1.9(1)$. Instead, $\sigma_M = 0.34$ appeared to be not only much less than $\sigma_V$ but even lower than that for nanoparticle cores, $\sigma_v = 3\sigma_d = 1.15$, with $\sigma_d$ from Table 1 (left column). This discrepancy is suggested to arise from the finite distance at which the nanoparticle magnetic moments correlate inside an aggregate. From Monte-Carlo simulations for growth kinetics of magnetic nanoclusters with d-d

**Table 3**
Parameters obtained from M2 measurements and identified with aggregates: without account of magnetic correlations (left column) and with account of finite-radius magnetic correlations (right column).

| | Infinite correlation radius | Finite correlation radius |
|---|---|---|
| Saturation magnetization $M_s$ , emu/mole(Fe) | 257.5(1.6) | 702(1) |
| Saturation magnetization per Fe-ion$\langle\mu\rangle$ , $\mu_B$ | 0.0454(3) | 0.127 |
| Median magnetic moment $M_0$ , $\mu_B$ | 45240(160) | |
| Mean magnetic moment $\bar{M}$ , $\mu_B$ | 47950(170) | 61200(100) |
| Distribution width $\sigma_V$ ($\sigma_M$) | 0.341(2) | 2.11 |
| Mean anisotropy energy $E_a$ , K | 171(3) | 254(22) |
| Anisotropy field $H_a$ , Oe | 53.2(11) | 61.8(7) |
| Anisotropy axis direction $\Psi$ , deg. | 10.71(17) | 12.7(1) |
| Damping constant $\alpha$ | 0.2246(8) | 0.2283(7) |
| Neél time $\bar{\tau}_N$ , s | 1.442(7) $\cdot 10^{-9}$ | 1.81(1) $\cdot 10^{-9}$ |
| Longitudinal relaxation time $\tau_\parallel$, s | 1.832(9) $\cdot 10^{-9}$ | 2.60(1) $\cdot 10^{-9}$ |
| Mean volume $\bar{V}$, nm$^3$ | 2.565 $\cdot 10^4$ | 3.80 $\cdot 10^4$ |
| Mean diameter $\tilde{D}$ , nm | 36.6(2) | 41.7 |
| Mean diameter $\bar{D}$ , nm | 36.1(2) | 25.4 |
| Mean number of Fe ions $\bar{N}$ | 1.04 $\cdot 10^6$ | 1.54 $\cdot 10^6$ |

coupling [11], magnetic interparticle correlations decay with increasing the distance and vanish within a finite correlation radius. A rigorous account of magnetic correlations in the nonlinear response is a complicated problem. Instead, a cutoff function was introduced for the aggregate magnetic moment *imitating* the real correlations:

$$g(r) = \begin{cases} 1, & r \leq r_0 = r_c \exp(\lambda^{-1}) \\ 1 - \lambda\ln\dfrac{r}{r_0}, & r_0 < r < r_c \\ 0, & r > r_c \end{cases} \quad (7)$$

where $r$ is the distance from the center of the aggregate. Its linear dependence on $-\ln r$ fairly well approximates the simulated correlation function for the distances not too close to the correlation radius $r_c$

[11]. The cutoff function modulates the density of magnetic moment inside the aggregate. Averaging the magnetic moment over the aggregate volume leads to correction of the former by the factor

$$K(V) = 1 - \frac{\lambda}{3}\left(\ln\frac{V}{V_0} + \frac{V_0}{V} - 1\right)$$

for the aggregate volume $V$ in the region $V_0 < V < V_c$ ($2r_0 < D < 2r_c$), where $V_0 = 4\pi r_0^3/3$ and $V_c = 4\pi r_c^3/3$. For $V < V_0$ ($D < 2r_0$) $K = 1$, and $K = 0$ if $V > V_c = 4\pi r_c^3/3$ ($D > 2r_c$) assuming no magnetic correlations in very large aggregates. With the cutoff function introduced, the magnetic moment distribution is no more lognormal, whereas the volume distribution is still assumed lognormal.

The results of the data treatment with account of the cutoff and with $\lambda$ and $r_0$ also as variable parameters are presented in Table 3 (right column). Parameters of the cutoff function are $\lambda = 0.283$ and $r_0 = 1.00$ nm. The latter value and the correlation radius $r_c = 34.2$ nm should be further corrected by a certain factor $\xi > 1$ (see Subsection 4.3.3).

The magnetic correlations account is seen to noticeably modify most of the parameters. First of all, the volume distribution width, $\sigma_V = 2.1$, appreciably increased and became quite close to this for aggregates observed in microscopy scans. The mean magnetic moment $\bar{M}$ raised by 28%. The intensive quantity $M_s$ after correction became almost three times larger, 702 vs. 258 emu/mole(Fe), as the former value belongs only to the magnetically correlated region while the latter one is the mean over-aggregate magnetization.

Thus, magnetic moments of SPIONs in aggregates are, to a great extent, disordered. The magnetic-moment and size (or volume) distributions width of the aggregates evaluated by various techniques based on magnetic measurements can be appreciably underestimated if finite-radius magnetic correlations are disregarded. Besides, considerable magnetic disorder due to anisotropic character of the interparticle dipolar forces appreciably diminishes the total aggregate magnetic moment. Being interpreted at simple manner, this could lead to the spurious conclusion on, e.g., a dimer structure of the aggregates.

*4.3.2. Quantifying magnetization dynamics and magnetic anisotropy of aggregates*

The damping constant defines the magnetic moment dissipation rate. It is the most robust fit parameter insignificantly correlating with the others and well reproduced even with no correlations account. High susceptiveness to magnetization dynamics is the main virtue of M2 technique. The present small value $\alpha \sim 0.2$ highlights the precession terms in Eqs. (2b) and (2c) pointing out an appreciable role of precession in relaxation of the magnetic moment. In the opposite, overdamped, case $\alpha \geq 1$, the relaxation would have been of purely thermal diffusion type.

The longitudinal relaxation times in the left and right columns of Table 3 were calculated with the interpolated values $Q \cong 1.270$ and $1.434$ corresponding to $\beta E_a \cong 0.5777$ and $0.857$ [34], respectively.

The aggregates were found to be magnetoanisotropic. However, the obtained "easy axis" anisotropy energy $E_a = 254$ K implies an unexpectedly small conventionally defined blocking temperature with the value $T_B = E_a/25 \cong 10$K more inherent to *noninteracting* magnetic nanoparticles [11]. Notice, the saturation magnetization per Fe ion $\langle\mu\rangle \sim 10^{-1}\mu_B$ is one order smaller than the magnetic moment of the magnetite Fe ion $\mu \sim 1\mu_B$. This finding is indicative of strong disorder of the SPION magnetic moments inside an aggregate. Orientation of SPIONs anisotropy axes during formation and growth of the aggregate is governed mainly by d-d coupling between the magnetic moments, the strength and the sign depending on the mutual position of the interacting nanoparticles. As a result, anisotropy axes and magnetic moments of nanoparticles inside a large aggregate are oriented almost randomly, as evidenced also by Monte-Carlo simulations [11].

In the different approach, randomly positioned dipoles tend to align due to interaction with the mean dipolar field [35]. However, the position randomness leads to fluctuations inhibiting the ordering. At low densities of dipoles, fluctuations dominate preventing the ordering, whereas at high densities, the mean field dominates and the ordering is possible. Thus, the partial magnetic ordering is a compromise between the two opposite tendencies.

The observed relatively small mean magnetic moment of the aggregates, only twice as large as the nanoparticle core magnetic moment (estimated in Subsection 5.2), is a result of incomplete mutual compensation of the moments correlated by dipolar forces sufficiently strong at the measurement temperature (also in Subsection 5.2).

The angle $\Psi$ indicates the predominant orientation of the aggregates anisotropy axes in the colloid relative to the applied magnetic field. In the isotropic case, when the aggregates anisotropy axes are completely disordered, this angle roughly corresponds to the cosine square averaged over $4\pi$, in the way anisotropy enters the magnetic potential $\mathcal{H}$, viz., $\Psi_{eff} = \cos^{-1}[(\overline{\cos^2\Psi})^{1/2}] \cong 55°$. The obtained value 12.7° is too small to agree with this assumption. The small $\Psi$ value is suggested to result from the orienting effect of the external magnetic field which tends to align the aggregates. The mean energy of the aggregate magnetic moment in the steady field, $\bar{E}_H = \bar{M}H$, exceeds the thermal energy in the great part of the measured $H$-region. Thus, in the field $H = 100$ Oe, $\bar{E}_H = 410\ K$. The anisotropy energy compared to the measurement temperature is also large enough to ensure the sufficient coupling between magnetic and rotational degrees of freedom. The moderate Boltzmann factor $\exp(-\beta E_a) \cong 0.42$ characterizes the effect of thermal disordering on magnetic moment with respect to the anisotropy axis.

To verify this suggestion, first, the same M2 data were once more treated assuming random distribution of the anisotropy axes, i.e. the function in Eq. (6) was additionally averaged over the axes directions using the Gauss quadrature [36]. The resultant fit quality turned out to be noticeably worse, with $\chi^2$ three times greater. Second, an additional M2 measurement was performed for the frozen colloidal solution, at the temperature 260 K, after zero-field cooling to retain the orientation disorder. The obtained value $\Psi \cong 46°$ turned out to be fairly comparable with $\Psi_{eff}$, despite rather an approximate character of the $\Psi_{eff}$ estimation. Third, in aqueous colloids, an aggregate experiences Brownian diffusion with the relaxation time $\tau_B = 3\beta\eta V_h$, where $\eta = 0.87$ mPa·s is the viscosity of water at the measurement temperature 297 K, and $V_h = \pi D_h^3/6 \cong 7.4 \cdot 10^4$ nm³ is the hydrodynamic volume estimated with $D_h = 52$ nm from Table 2 (left column). The Brownian relaxation time $\tau_B \cong 4.8 \cdot 10^{-5}$ s appeared to be much less than the round-up cycle of the scan field, 0.125 s, thus, meeting the adiabatic condition. On the contrary, $\tau_B \gg 1/f \cong 6.4 \cdot 10^{-8}$ s and the rotational degrees of freedom are unaffected by the high-frequency ac magnetic field. Besides, as $\tau_B \gg \tau_\| \sim 10^{-9}$ s, its contribution to the effective relaxation time $\tau_{eff}$ defined via the relation $\tau_{eff}^{-1} = \tau_\|^{-1} + \tau_B^{-1}$ is negligibly small.

All estimations ensure the suggestion on orienting the aggregates by magnetic field in the liquid colloidal solution. The orientational effect was observed in the similar object by nonlinear Faraday rotation even in smaller magnetic fields, $H \leq 40$ Oe [14]. Orientational mobility of magnetic nanoparticles in suspensions under magnetic field has been studied also by EMR techniques [37-39].

As the recovered volume distribution obtained for the aqueous colloidal solution satisfactorily fits the distribution observed by TEM on the freeze-dried samples, no noticeable additional aggregation occurs, most probably, when drying the solution to obtain the TEM specimen.

*4.3.3. Distinguishing between magnetic and nonmagnetic components of the colloid*

The recovered volume distributions, all normalized by unity in maxima, are presented in Fig. 5. The narrow dashed-line peak centered at $10^4$ nm³ corresponds to the treatment without account of correlations while the thick-line broad peak centered at $2 \cdot 10^2$ nm³ is a recovery with the correlations account.

The TEM histograms for nanoparticle cores and SPION aggregates (Fig. 2) recalculated to the volume distributions are also presented by open and filled circles, respectively, together with their best fits (respective solid curves). The cutoff function (Eq. (7)) monotonously going down with the volume increase (red dotted line) is shown, as well. The recovered volume distribution (thick line), being of almost the same width as the aggregates fit, is shifted to lower volumes. This misfit is due to the fact that the aggregates distribution belongs to the whole colloid magnetite+dextran whereas the recovered distribution concerns only the magnetite fraction. With the known parameters for both the distributions, it is possible to estimate the dextran shell thickness of nanoparticles constituting the aggregates.

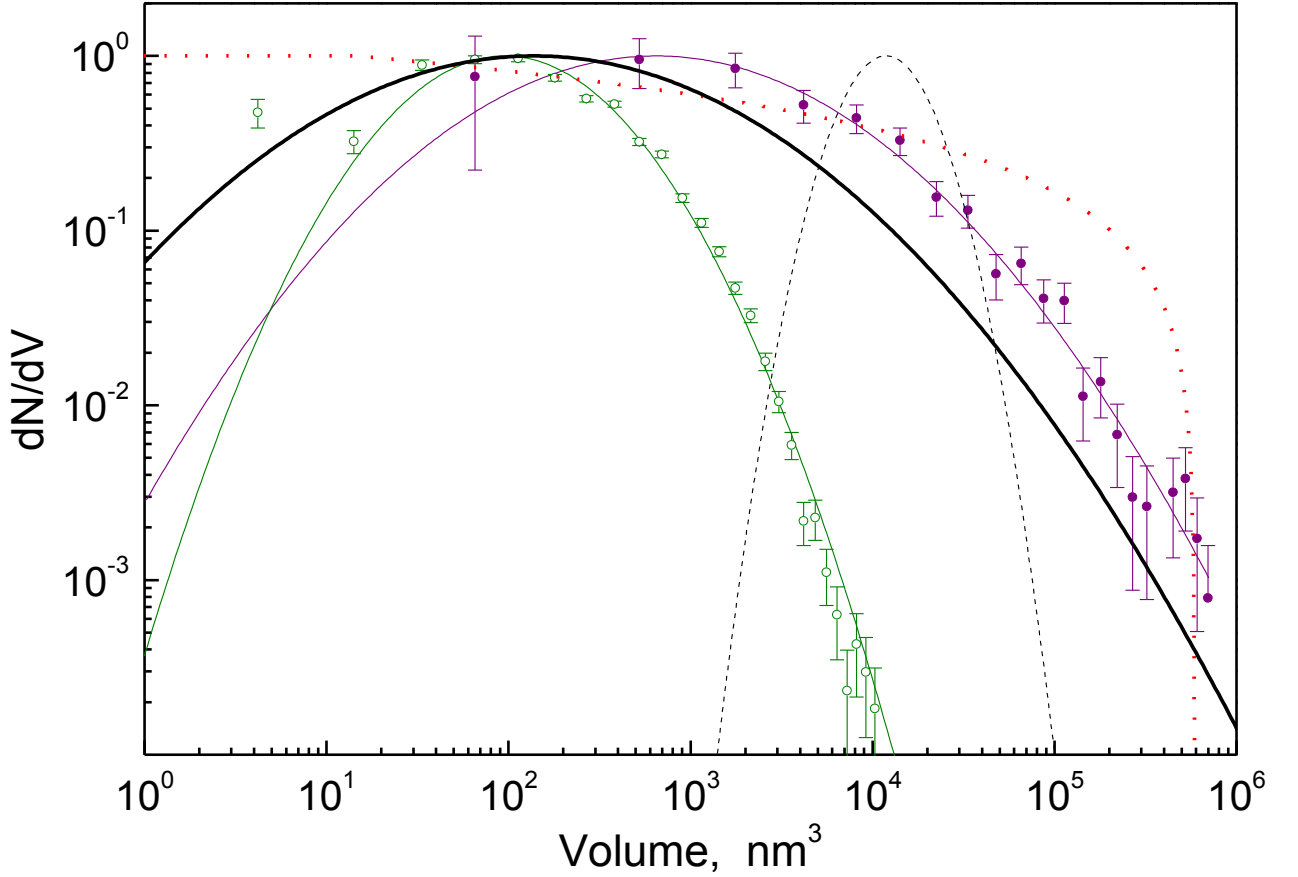

**Fig. 5.** Volume distributions, all normalized by unity in maxima: (i) distribution of SPIONs from TEM centered at $10^2$ nm$^3$ (green open circles) and its best fit (solid line), (ii) distribution of aggregates from TEM centered at $8 \cdot 10^2$ nm$^3$ (purple filled circles) and its best fit (solid line), (iii) distribution recovered from M2 data fit (narrow dashed peak centered at $10^4$ nm$^3$), (iv) distribution recovered from M2 data fit with correction on magnetic correlations (see Subsection 4.3.1) (black thick curve), (v) cutoff function imitating interparticle magnetic correlations with singularity at $6 \cdot 10^5$ nm$^3$ corresponding to the correlation radius (red dotted line).

From the recovered mean volume $\bar{V}$ (Table 3, right column) and the TEM mean nanoparticle-core volume $\bar{v} = 688(35)$ nm$^3$, one obtains the mean number of SPIONs in the aggregates, $n_c = \bar{V}/\bar{v} \cong 55$. The nanoparticle specific volume (the average volume per one nanoparticle in the aggregate) is obtained from the TEM mean aggregate volume $V_c = 1.33 \cdot 10^5$ nm$^3$ as $v_c = V_c/n_c = 2420$ nm$^3$. After that, the dextran shell thickness of the nanoparticle can be determined as

$$\delta = \frac{1}{2}(d_c - \tilde{d}).$$

Here, $d_c$ is the mean diameter of SPIONs related to the mean volume per nanoparticle $v_c$ as $\varphi v_c = \pi d_c^3/6$ where $\varphi$ is the occupation factor arising due to the empty space between nanoparticles in the aggregate. Its value is $\varphi = 0.56$ and $0.64$ for friable and compact irregular packing, respectively, resulting in $d_c = 13.7$ nm for the former and 14.4 nm for the latter case, respectively. The mean core diameter $\tilde{d} = 11.0$ nm relates to the mean volume $\bar{v}$ as $\bar{v} = \pi \tilde{d}_c^3/6$. As a result, the dextran shell width is estimated to lie in the reasonable interval $\delta = 1.4 - 1.7$ nm where the lower number corresponds to the friable irregular packing and the upper to the compact one.

Comparing the mean (magnetite only) volume of aggregates $\bar{V}$ obtained from M2 and the mean volume $V_c$ obtained from TEM, one immediately estimates the portion of the volume magnetite occupies in aggregates, $x = \bar{V}/V_c \cong 0.28$. As magnetite cores are uniformly spread over the aggregate, the cutoff function parameters $r_0$ and $r_c$ obtained from the M2 data treatment should be rescaled by the factor $\xi = 1/\sqrt[3]{x} \cong 1.52$ yielding the true values $r_0 \cong 1.52$ nm and $r_c \cong 51.9$ nm. The correlation radius $r_c$ noticeably exceeds the mean aggregate radius 21.5 nm (Table 1, right column) whereas the size $r_0$ of the complete-correlation area is even smaller than the mean magnetite core radius, 4.7 nm (Table 1, left column). The latter is a consequence of strong anisotropy of the d-d coupling and a statistical character of the quantity $r_0$.

*4.3.4. Complementary remarks*

From Monte-Carlo simulations [11], the growth kinetics of aggregates exhibits the tendency to fractal formation. The fractal dimension $\nu$ is conventionally defined by the relation $n \propto D^\nu$ where $n$ is the number of nanoparticles constituting an aggregate of the size $D$. From TEM, the aggregate size has the lognormal distribution with the width $\sigma_D = 0.63$ (Table 1) while the number of nanoparticles $n = V/\bar{v}$ obtained from M2 also has the lognormal distribution with the width $\sigma_V = 2.1$ (Table 3). With Eq. (1), the fractal dimension can be evaluated as $\nu = \sigma_V/\sigma_D \cong 3$. Thus, no fractal formation tendency is observed in the colloidal system under study. The aggregates consisting of ~ 10 – 100 nanoparticles are composed compactly and still too small even for a hint on the fractal structure.

Thus, the data treatment with no correction on magnetic correlations, instead of computer time consuming correlations account, may be implemented only for rough quantifying.

Note also that no noticeable difference of M2 parameters was found for colloidal solutions with the concentrations 0.02 and 2 mM(Fe)/L. This finding indicates stability of the aggregates in the concentration range examined.

## 5. Electron magnetic resonance

*5.1. Measurements and data treatment*

The measurements performed were supplemented with EMR data obtained from the liquid colloid with the concentration 4 mM(Fe)/L. The EMR spectra were recorded with the special homemade X-range spectrometer operating at the frequency $F = \omega/2\pi = 8.54$ GHz, which provided high sensitivity at registration of wide resonance lines [40]. The spectrometer was supplied by the cylindrical two-mode balanced cavity with $TE_{111}$-type of electromagnetic oscillations. The steady magnetic field **H** was directed along the cylinder axis $z$. The sample was placed at the bottom of the cavity where it was affected by the linearly polarized ac field **h** directed along the $x$-axis perpendicular to **H** (excitation $xz$ plane). The detection $yz$ plane was perpendicular to the excitation one and, thus, the detected signal was proportional to the gyrotropic (off-diagonal) component of the susceptibility tensor, $\chi_{yx}(\omega)$, corresponding to the $y$-component of the induced magnetic moment $M_y(\omega) = \chi_{yx}(\omega)h_x(\omega)$. The deep frequency-independent uncoupling between the excitation and detection modes made it possible to use a microwave source with the high oscillation power (~1 W) without frequency- or amplitude noise at the input of the detector, thus, providing high spectrometer sensitivity. This facility has proved its efficiency in a number of condensed matter studies [17, 41, 42].

The EMR signals proportional to the mixture of the dispersion $\chi'_{yx}$- and absorption $\chi''_{yx}$ parts of the magnetic susceptibility $\chi = \chi' - i\chi''$ were registered as functions of the magnetic field ranging from 260 to 6400 Oe.

The spectrum, measured at the temperature 285 K, is presented in Fig. 6 (circles).

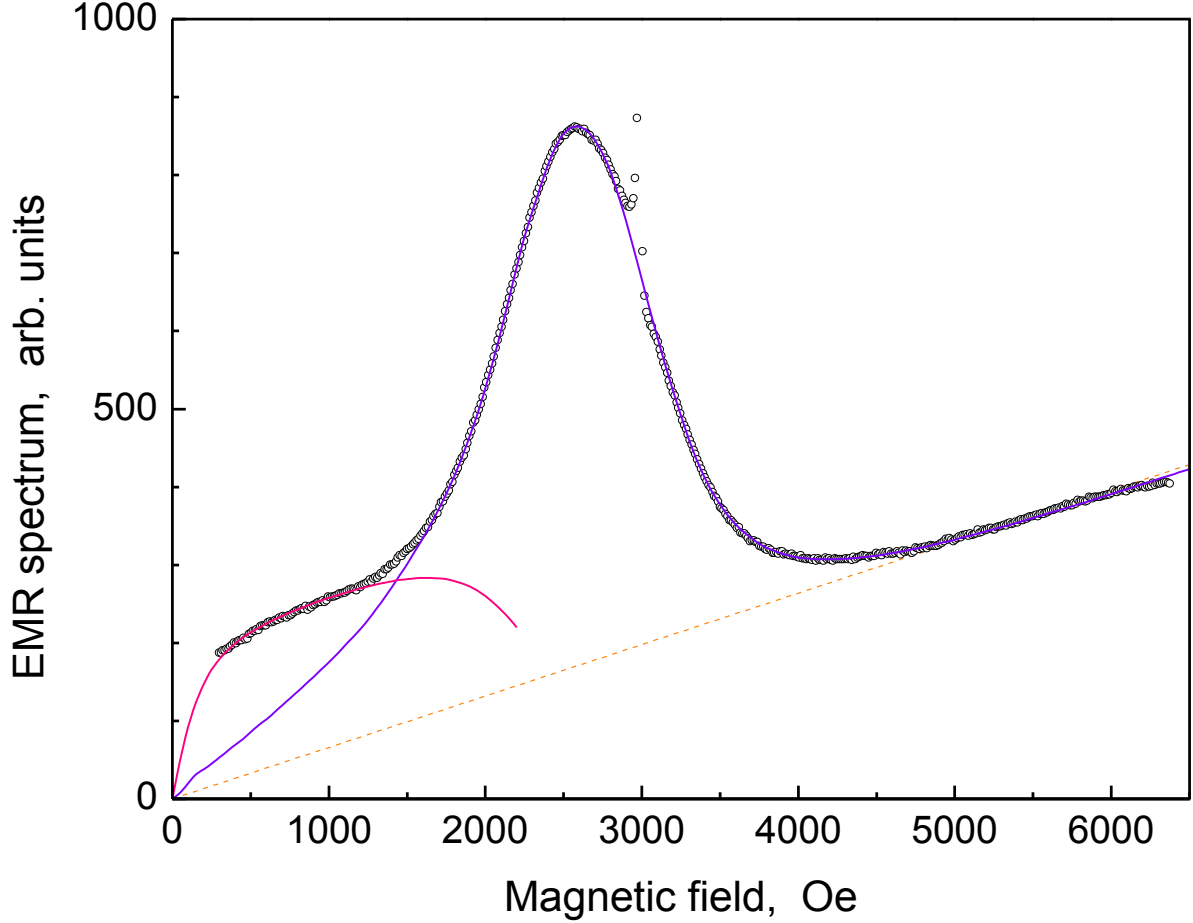

**Fig. 6.** EMR spectrum (circles): sharp peak at 3 kOe is nitroxyl radical signal used as a calibration witness; solid curves are best fits: for magnetically correlated (H<1250 Oe) and independent (H>1650 Oe) particles (see Subsection 5.2) with crossover at 1430 Oe; straight dashed line is background Hall signal proportional to magnetic field.

The sharp peak centered at 3 kOe comes from nitroxyl radicals used as a calibration witness. The main signal was treated with the similar GLL formalism as used for treatment of the M2 data assuming the SP behavior to persist also at the EMR frequency $F$. Instead of Eq. (5) valid for the diagonal response [43], i.e. parallel to the excitation field, the non-diagonal induced moment implies the form:

$$m_k(\omega) = \sqrt{\frac{2\pi}{3}}\left[(\gamma'_2 + i\gamma'_1)c^k_{11}(\omega) - (\gamma'_2 - i\gamma'_1)c^k_{1-1}(\omega)\right]. \tag{8}$$

The function fitting the EMR response, with $k = 1$ corresponding to the linear susceptibility, reads:

$$M_1(H) = M_s \int (m'_1 \sin\Theta + m''_1 \cos\Theta) f_M \, dM \tag{9}$$

where $\Theta$ is the angle mixing the real $m'_1$- and imaginary $m''_1$ parts of the induced moment given by Eq. (8). To fit the spectrum, Eq. (9) was accompanied by the background Hall signal proportional to the magnetic field $H$ coming from the cavity material of the spectrometer [40].

## 5.2. Aggregates vs. independent nanoparticles

As seen in Fig. 6, two field regions exist where the signal is fitted in different ways. The best-fit curve well describes the peak and the high-field "tail" of the signal in the region $H > H_2 = 1650$ Oe, whereas its extrapolation to lower fields strongly deviates from the measured spectrum. This failure is caused by competition between the interparticle d-d coupling and interaction of nanoparticles with the external magnetic field. In the low-field region $H < H_1 = 1250$ Oe, magnetic moments of SPIONs constituting the aggregates are coupled by dipolar forces and the EMR signal is a response of the aggregates within the correlation radius as it occurred in M2 measurements. At elevated magnetic fields, the d-d coupling is broken and the signal becomes an additive response of independent nanoparticles. In Fig. 6, the signal in the lower-field region is well described by the same Eq. (9) with the aggregate parameters determined from M2 measurements (Table 3), the only variable parameters being the normalization factor and the mixing angle. Both the curves intersect at the crossover field $H_c = 1430$ Oe, while at zero field, they fall to zero. The characteristic interparticle dipolar energy reads

$$E_d = 4\pi \frac{\bar{m}^2}{v_c}.$$

Recall that $v_c = 2950$ nm$^3$ is the mean volume per nanoparticle inside an aggregate, $\bar{m} = \mu \bar{n}$ is the mean magnetic moment of nanoparticles where $\mu$ is the Fe-ion magnetic moment averaged over the unit cell and $\bar{n} = \bar{v}/v_0 \cong 27900$ is the mean number of Fe ions per nanoparticle. The mean energy of the nanoparticle magnetic moment in the field $H$ is, merely, $E_H = \bar{m}H$. The crossover field corresponds to the condition $E_d = E_H$:

$$H_c = \frac{4\pi \bar{m}}{v_c}.$$

From the resonance line shape, in addition limited by the crossover, not all SP parameters can be reliably evaluated. In particular, mainly due to large uncertainty in calibration of the signal, the nanoparticle magnetic moment is hardly available from the present EMR data. This forces to turn to off-site magnetization measurements.

The saturation magnetization of magnetite nanoparticles revealed dependence on size, shape, manufacturing conditions, and matrix- or coating material, if exists [27, 44, 45]. Spherical nanoparticles ~10 nm in diameter were found to become saturated in strong magnetic fields at the temperature 300 K up to the magnetizations ~60 - 80 emu/g [27, 44, 45]. These values correspond to the Fe-ion magnetic moment $\mu \cong 0.8 - 1.11 \mu_B$ resulting in $\bar{m} \cong (2.2 - 3.10) \cdot 10^4 \mu_B$. With these numbers, the crossover field falls into the interval $1120 - 1490$ Oe. Thus, the suggested explanation of the EMR spectrum by interplay of d-d coupling and the external magnetic field seems to be highly credible.

With increasing the field over $H_c$, the SP dynamics dramatically changes. The damping constant reduces from $\alpha = 0.228$ to $0.125$ while the Néel time $\bar{\tau}_N$ and the longitudinal relaxation time $\tau_\parallel$ increase by five times. This effect, visualized here by EMR, should be directly observed in magnetorelaxation measurements. Its account in practical applications of SPION colloids would be important.

The organic coating may somewhat diminish the magnetization of SPIONs [27, 45] while the measured $H_c$ is quite close to the upper limit corresponding to the magnetic moment of uncoated nanoparticles [44, 45]. This might imply only a weak effect of the dextran coating on the magnetization of SPIONs under study, if any.

By the way, the crossover field is much greater than the maximal field in M2 measurements, $H_m = 300$ Oe, and the respective mean Zeeman energy for independent nanoparticles, $E_H \cong 500$ K, is much less than $E_d \cong (1.7 - 3.1) \cdot 10^3$ K. This explains the correlated character of M2 response for the whole array of magnetic moments.

The large dipolar energy much exceeding the measurement temperature ensures also the dipolar-glass type of the aggregate magnetic structure.

The large effective value of anisotropy axes directions $\Psi \cong 44°$ obtained from EMR is indicative of the great extent of disorder of their orientations inside the aggregates, similarly to the case of frozen colloid in the M2 measurements mentioned above. The respective anisotropy energy with $\beta E_a \cong 1.1$ seems to be large enough to ensure orienting an isolated nanoparticle by the steady magnetic field. However, orientation disorder inside the aggregate cannot be eliminated even by rather high EMR magnetic field well exceeding the anisotropy field $H_a \cong 700$ Oe, due to chemical bonds between the SPIONs combined in aggregates.

The distribution width of the SPIONs magnetic moments $\sigma_m = 1.09$ is quite close to the expected value $\sigma_v = 3\sigma_d \cong 1.15$ obtained from TEM for nanoparticle cores (Table 1).

## 6. Conclusions

Second-harmonic magnetic response to a weak ac field was employed to study an aqueous colloidal solution of dextran-coated magnetite nanoparticles applicable in biomedicine. Magnetic field dependences of real and imaginary parts of the response in conjunction contain full information on magnetic properties of the colloid ferrofluid. Expectedly, the magnetic colloid has the tendency to form clusters. The data processing based on the stochastic Gilbert-Landau-Lifshitz equation was applied to describe the system via a set of parameters, including the mean magnetic moment of the aggregates, the damping constant, the longitudinal relaxation time, the magnetic anisotropy field and energy, and others.

With M2 technique, magnetic correlations inside the aggregates arising from dipole-dipole coupling were distinguished. Their account in the data treatment enabled to recover the magnetic correlation radius, the size and volume distributions of aggregates, the concentration of aggregates in the solution, the mean magnetic moment per nanoparticle and the energy of dipole-dipole interaction between nanoparticles. The obtained parameters add to and agree with the data obtained from transmission electron microscopy, dynamic light scattering and electron magnetic resonance. In particular, combined analysis of the M2 and TEM data enabled to distinguish between magnetic and nonmagnetic components of the colloid.

To recover correctly the magnetic-moment and volume distributions by techniques based on magnetic measurements, account of finite-radius magnetic correlations in aggregates is essentially required.

The aggregates were argued to have dipolar-glass-type structure and possess magnetic anisotropy. The anisotropy turned out to be strong enough to ensure coupling of magnetic and rotational degrees of freedom resulting in noticeable orienting the aggregates by the steady magnetic field of the order $10^2$ Oe.

EMR spectra are well described in the framework of superparamagnetic dynamics. Herewith, the lower-field part of the signal is generated by magnetically correlated aggregated nanoparticles, similarly to the case of nonlinear magnetic response, whereas at higher fields, the signal is formed by independently responding nanoparticles due to break of interparticle dipole-dipole coupling by the external magnetic field. The break is accompanied with abrupt acceleration of superparamagnetic dynamics in increasing magnetic field. This effect should be accounted for and can be used in practical applications of SPION colloids. For instance, probing magnetization dynamics of SPIONs accumulated in tissues by magnetorelaxometry or any other suitable technique may specify whether the MNPs are in aggregated or nonaggregated state.

The present research exemplifies application of the novel M2-based procedure for comprehensive quantitative characterization of a wide variety of SP systems, particularly, directed toward biomedical applications. Particular data obtained in this study will be employed to examine the state of functionalized SPIONs accumulating in tissues.

## Acknowledgments

The authors are grateful to the management of Peter the Great Saint-Petersburg Polytechnic University Supercomputing Center for making available the computational resources, as well as to A. V. Arutyunyan for helpful discussions and A. M. Ischenko for permanent interest and support of the study.

This research did not receive any specific grant from funding agencies in the public, commercial, or not-for-profit sectors.